\documentclass[10pt,prl,twocolumn,showpacs,superscriptaddress,floatfix]{revtex4-2}
\usepackage{graphicx}
\usepackage{amssymb}
\usepackage{amsmath}
\usepackage{amsthm}
\usepackage{bm}
\usepackage[colorlinks,
linkcolor=blue,      
anchorcolor=blue, 
citecolor=purple]{hyperref}
\usepackage{color,xcolor}
\usepackage{subfigure}
\usepackage{algpseudocode}
\usepackage{soul}
\usepackage{algorithm}
\usepackage{algorithmicx}
\usepackage{lineno}
\usepackage{soul,xcolor}
\usepackage{color}

\usepackage{graphicx}
\usepackage{dcolumn}
\usepackage{bm}
\usepackage{graphicx}
\usepackage{textcomp,mathcomp}
\usepackage{enumerate}

\usepackage{lipsum} 

\floatname{algorithm}{Algorithm} 

\begin{document}

\preprint{APS/123-QED}

\title{Learning to Restore Heisenberg Limit in Noisy Quantum Sensing via Quantum Digital Twin}

\author{Hang Xu}
\thanks{Hang Xu and Tailong Xiao contributed equally.}
\affiliation{State Key Laboratory of Photonics and Communications, Institute for Quantum Sensing and Information Processing, Shanghai Jiao Tong University, Shanghai 200240, People’s Republic of China}%

 \author{Tailong Xiao}%
\email{tailong\_shaw@sjtu.edu.cn}
\affiliation{State Key Laboratory of Photonics and Communications, Institute for Quantum Sensing and Information Processing, Shanghai Jiao Tong University, Shanghai 200240, People’s Republic of China}%
 \affiliation{Hefei National Laboratory, Hefei, 230088, People’s Republic of China}
\affiliation{Shanghai Research Center for Quantum Sciences, Shanghai, 201315, People’s Republic of China}

 \author{Jingzheng Huang}%
\affiliation{State Key Laboratory of Photonics and Communications, Institute for Quantum Sensing and Information Processing, Shanghai Jiao Tong University, Shanghai 200240, People’s Republic of China}%
 \affiliation{Hefei National Laboratory, Hefei, 230088, People’s Republic of China}
\affiliation{Shanghai Research Center for Quantum Sciences, Shanghai, 201315, People’s Republic of China}


\author{Jianping Fan}
\affiliation{AI Lab, Lenovo Research, Beijing 100094, People’s Republic of China}%

 \author{Guihua Zeng}%
 \email{ghzeng@sjtu.edu.cn}
\affiliation{State Key Laboratory of Photonics and Communications, Institute for Quantum Sensing and Information Processing, Shanghai Jiao Tong University, Shanghai 200240, People’s Republic of China}%
 \affiliation{Hefei National Laboratory, Hefei, 230088, People’s Republic of China}
\affiliation{Shanghai Research Center for Quantum Sciences, Shanghai, 201315, People’s Republic of China}

\date{\today}

\begin{abstract}
Quantum sensors leverage nonclassical resources to achieve sensing precision at the Heisenberg limit, surpassing the standard quantum limit attainable through classical strategies.
However, a critical issue is that the environmental noise induces rapid decoherence, fundamentally limiting the realizability of the Heisenberg limit. 
In this Letter, we propose a quantum digital twin protocol to overcome this issue. 
The protocol first establishes observable-constrained state reconstruction to infer random errors in the decoherence process, and then utilizes reinforcement learning to derive adaptive compensatory control strategies.
Demonstrated across discrete, continuous variable and multi-qubit circuit systems, our approach bypasses quantum state tomography's exponential overhead and discovers optimal control schemes to restore the Heisenberg limit.
Unlike quantum error correction or mitigation schemes requiring precise noise characterization and ancillary qubits, our autonomous protocol achieves noise-resilient sensing through environment-adaptive control sequencing. 
This work establishes quantum digital twin protocol as a generic methodology for quantum control, proposing a noise-immune paradigm for next-generation quantum sensors compatible with NISQ-era experimental constraints.
\end{abstract}

\maketitle


\textit{Introduction.}---Quantum sensing aims to achieve high-precision parameter estimation by utilizing quantum resources.
While classical approaches are constrained by the standard quantum limit (SQL), quantum strategies employing coherence and entanglement can achieve the Heisenberg limit (HL) \cite{HL1,HL2,HL4,HL5}.
Recent advancements have further demonstrated that unconventional quantum resources including indefinite causal order \cite{ico1,ico2,ico3,ico4,ico5} and criticality-enhanced quantum phase transitions \cite{critcal1,critical2,critcal3,critcal4} may enable precision scaling beyond even the HL under ideal conditions.
However, the Heisenberg limit is usually unattainable considering noise and decoherence \cite{noisysensing0,noisysensing1,noisysensing2,noisysensing3,noisysensing4,noisysensing5}, which is known as the no-go theorem   \cite{nogo} of quantum metrology.

To address this problem, quantum error correction (QEC) protocols have been proposed to preserve quantum coherence through active stabilization techniques, potentially restoring HL-compatible precision in open quantum systems \cite{qec1,qec2,qec3,qec4,qec5,qec6,qec7,qec8}.
Nevertheless, their implementation requires ancillary qubits, precise system characterization, and strict satisfaction of error correction conditions \cite{qec5}. 
Meanwhile, quantum control methods provide partial noise mitigation \cite{control1,control2,control3,control4}, though fundamentally limited in restoring HL-scaling precision \cite{zss1}. These dual limitations are the core challenge for NISQ-era quantum sensing: circumventing noise induced no-go theorem to realize HL-compatible measurements under experimentally realistic conditions—particularly with unknown system dynamics \cite{inference1,inference2} and device imperfections \cite{nisq1}.

Digital twin \cite{digital1,digital2} is a virtual digital mapping of physical system that enable real-time simulation and predictive analytics through continuous data synchronization.
{While artificial intelligence has advanced quantum research \cite{ml1,ml2,ml3,ml4,ml5,ml6}, the digital twin paradigm remains largely unexplored in this context.
Inspired by neural-network quantum states \cite{nqs1,nqs2}, we define a quantum digital twin model as a classical model continuously updated from measurement records yet capable of capturing coherence, entanglement, and measurement backaction.
The quantum digital twin approach promises distinct advantages for NISQ devices, including adaptive control, dynamic error mitigation, and protocol optimization.}

In this Letter, we propose a quantum digital twin sensing (QDTS) protocol to overcome no-go theorem, operating through a closed-loop architecture comprising predictive modeling and decision optimization phases. 
In the prediction phase, we continuously monitor the environment of real system ${{\mathbb S}}$ to collect measurement signals \cite{monitor1,monitor2}.
Then we sample ${\mathbb S}$'s state and use the measurement signals to learn the features of ${\mathbb S}$ to build its digital twin model ${\tilde {\mathbb S}}$. 
The learned ${\tilde {\mathbb S}}$ can subsequently track ${\mathbb S}$ based on real-time measurement signals.
In the decision phase, the reinforcement learning (RL) agent directly extracts the features of ${\tilde {\mathbb S}}$ to learn how to control the real system ${\mathbb S}$ to maximize the sensing precision. 
The QDTS protocol in the prediction phase infers random errors from continuous measurement data across different measurement trajectories of the decoherent system, enabling real-time error compensation through RL in the decision phase. This eliminates decoherence and preserves unitary parameter encoding. 
In contrast, open-loop control methods (e.g., GRAPE \cite{grape1} and conventional RL \cite{control1}) cannot detect such errors and only apply uniform controls across different trajectories, offering transient decoherence mitigation.
Numerical verification across single-atom, quantum circuit, and continuous-variable platforms demonstrates HL recovery under unknown noise conditions.
Crucially, our protocol circumvents the density matrix reconstruction requirement that plagues most RL approaches in quantum control. We demonstrate that strategically selected observable expectations suffice for behavior characterization in quantum sensing applications. Our protocol leverages partial-observable extraction from measurement data rather than full quantum state tomography -- a methodology conceptually aligned with shadow tomography \cite{shadow1,shadow2,shadow3}.

\textit{Noisy quantum sensing.}---In quantum sensing, the estimation precision of a parameter $\omega$ is given by the variance ${\mathop{\rm Var}}(\hat \omega )$ of its estimator $\hat \omega$, and the lower bound of the variance is defined by the Cramér-Rao inequality \cite{cramer1,cramer2,qfi1}
\begin{equation}
{\mathop{\rm Var}}(\hat \omega ) \ge {M^{ - 1}}{\cal F}_C^{ - 1} \ge {M^{ - 1}}{\cal F}_Q^{ - 1},\label{inequality}
\end{equation}
where $M$ is the number of measurements. 
{The classical Fisher information (CFI) ${\cal F}_C$ is the upper bound on the precision of estimator $\hat \omega$ from measurement results under a specific measurement setup.
Optimizing the CFI over all possible measurement setups leads to Quantum Fisher Information (QFI) ${\cal F}_Q$.
For the purposes of specific measurements, we employ CFI as the indicator of sensing precision.}

In closed quantum systems without noise, realizing the HL is straightforward. 
For a frequency $\omega$ estimating with Hamiltonian $\omega {\sigma _z}$, the QFI of the system reaches $4{t^2}$ simply by free evolution. 
If expanded to a system of $N$ particles, the QFI reaches $4t^2N^2$ simply by setting the initial state to be a Greenberger-Horne-Zeilinger (GHZ) state \cite{ghz1}.
However, quantum systems are inherently open, with their quantum-enhanced sensing capabilities critically vulnerable to environmental decoherence. While achieving HL precision within coherence times, such systems exhibit noise-driven performance degradation below the SQL at extended timescales. 
The no-go theorem shows that the noisy sensing precision degrades rapidly on time scales if quantum error correction is not available. 
Besides, we often do not know the type and intensity of noise in the system or how it is coupled to the environment. This means that adaptive control and error correction strategies fail since we cannot explicitly model the dynamics of the system.

\textit{The QDTS protocol.}---Without loss of generality, we model the noisy system ${\mathbb S}$ as a form of its coupling to the environment ${\mathbb E}$:
\begin{equation}
{H_{SE}} = {H_S}(\theta) + g{H_I},
\label{noisyform}
\end{equation}
where ${H_S}(\theta)$ is the system's Hamiltonian with estimated parameters, $H_I$ is the coupling term between the system and the environment, and $g$ is the coupling or noise strength.
{We set $g=1$ in whole work.}
Here, the system and the environment evolve together as a whole. 
If the environment is ignored, the system undergoes non-Markovian noise evolution \cite{nmn1,nmn2,nmn3}. 
We adopt this form to describe the system evolution under non-Markovian noise. 
In case the time duration is highly small, it can be approximated as a Markov process, but we consider larger time scales.

\begin{figure}[htbp]
  \centering
\includegraphics[width=1\linewidth]{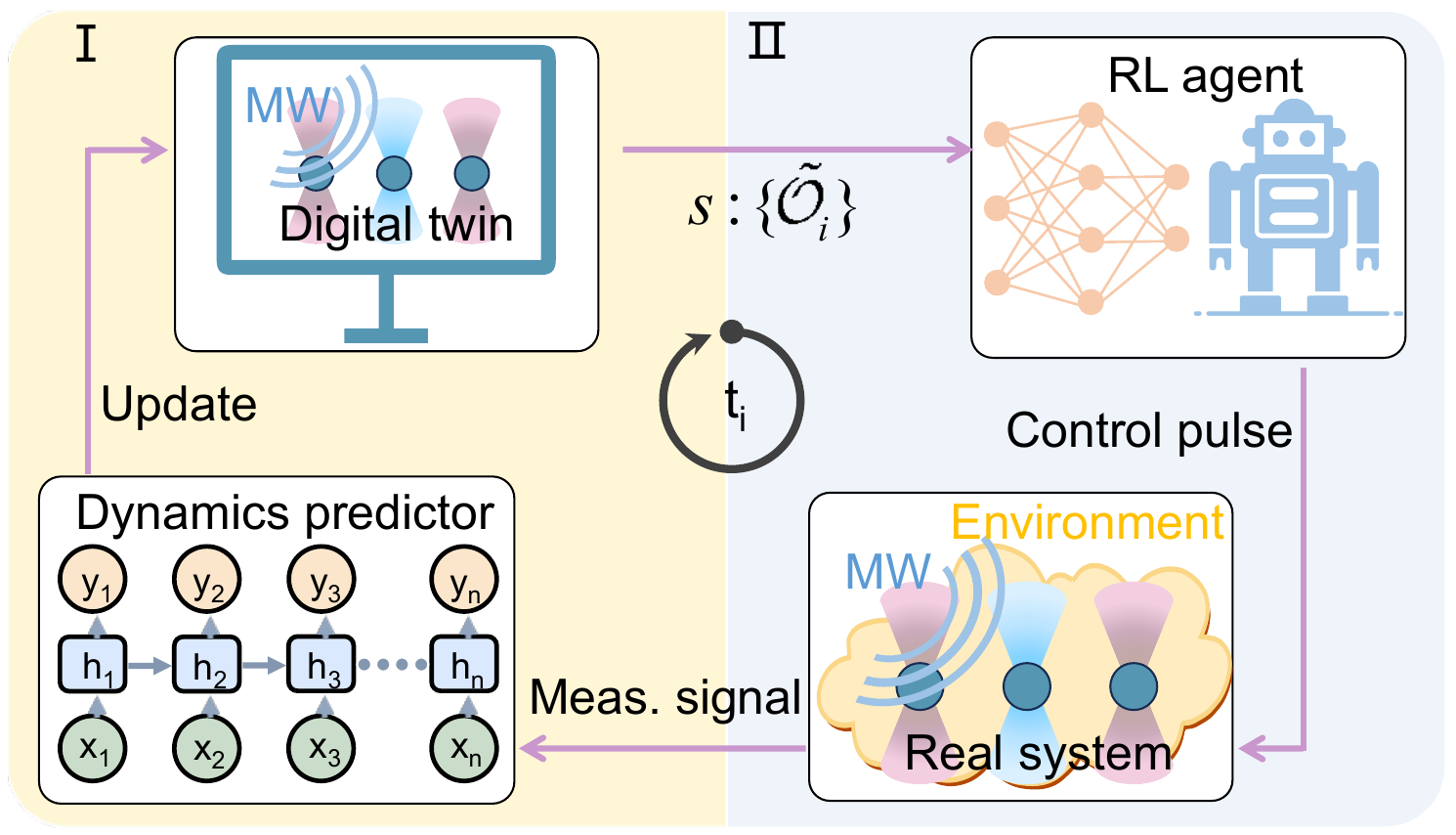}
  \caption{Schematic of the quantum digital twin sensing protocol. 
  The dynamics predictor can be recurrent neural networks or Transformer.}
  \label{frame}
\end{figure}

As shown in Fig.~\ref{frame}, the QDTS protocol consists of two phases: the prediction and decision phases. 
In the prediction phase, the prediction network generates digital twin model ${\tilde {\mathbb S}}$. The decision phase uses RL to control the real system ${\mathbb S}$, but the features as input of RL come from the twin. 
Once the prediction network is trained, the RL agent learns the optimal control strategy of the physical system and maximize the estimation precision of the unknown parameters.

 

In prediction phase, we first sample the initial state $S_0=\{{\left\langle {\cal O} \right\rangle _0}\}$ of the quantum system $\mathbb S$ and then let the system undergo free and controlled evolution. 
Meanwhile the environment is continuously monitored to collect continuous measurement signals ${\cal I}(t)$. 
Finally the system evolves to the final state $S_t=\{{\left\langle {\cal O} \right\rangle _t}\}$. The measurement signals are taken as input to a neural network to predict the expectations of observables via supervised training with $S_t$ as ground-truth.
The mean square error ${\cal L}(w) = \sum_\nu {{{( {{S_t^\nu} - {{\tilde S}^\nu_t}(w)})}^2}}$ is used to evaluate the prediction performance where ${\tilde S}^\nu_t (w)=F(\mathcal{I}^\nu(t))$ with $\nu$ denoting the $\nu$th trajectory, $F$ denoting the prediction network.
The trained network faithfully estimates the final state $\tilde S_t=\{ {\tilde{\left\langle {\cal O} \right\rangle} _t}\}$. We call ${\tilde {\mathbb S}}$ as the digital twin of real quantum system $\mathbb S$. 
Note that the prediction network only generates the digital twin model without characterizing the environment.
In this phase, the measurement signals are obtained by continuous monitoring the environment, which can be experimentally realized by photon counting or homodyne detection.
This nondemolition measurement preserves quantum coherence by avoiding wavefunction collapse, thereby maintaining uninterrupted parameter encoding dynamics for the unknown quantity $\theta$ throughout the sensing protocol.

In decision phase, continuous monitoring of the environment is still required.
The monitoring signals are fed into the prediction network to update the digtal twin such that it can track the evolution of the system in real time.
The RL agent can make decision with single-trajectory estimated twin state instead of collecting the measurement statistics with large number of repeated experiments.
At the $i$-th control cycle, the RL agent utilizes the digital twin model $\tilde{S}_{i}$ as input and then selects an control action $A$ from the predefined action space $\cal A$, where each $A$ corresponds to a specific unitary evolution operator that drives the physical system from state $\rho_i$ to $\rho_{i+1}$.
Meanwhile the prediction network receives quantum trajectories to update the twin's state to $\tilde{S}_{i+1}$. The RL agent computes the CFI as its input to the reward function $r_{i}= f({\cal F}_C)$. The goal of RL is to sequentially repeat the cycle $\mathcal{T}$ times to maximize the accumulated reward ${\cal R} = \sum\nolimits_{i = 1}^\mathcal{T} {{\eta ^i}{r_i}(\tilde{S}_i,A_i)}, $ where  $\eta$ is the discount rate. The policy network $\pi_{\zeta}(A|\tilde{S})$ parameterized by $\zeta$ is updated via gradient ascent to optimize the control strategy.

\begin{figure}[h]
  \centering
  \includegraphics[width=1\linewidth]{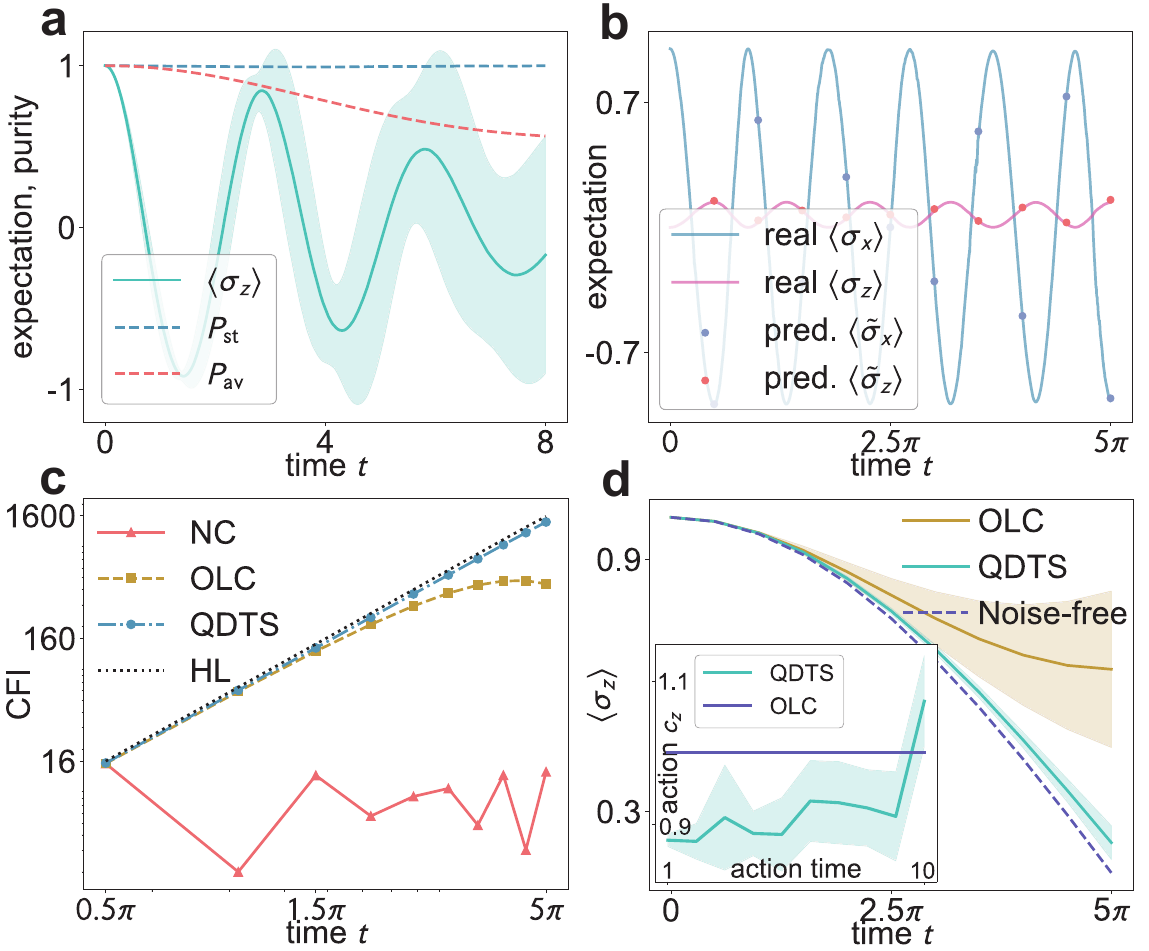}
  \caption{Results for the single atom system.
  (a) Dynamics and purity of the system under the continuous monitoring environment setup, where the shaded area indicates the standard deviation.
  $P_{\rm st}$ is the purity of the density matrix of a single trajectory, while $P_{\rm av}$ is the purity after averaging the density matrix over all trajectories.
  (b) Predictive performance of the digital twin model. 
  The solid lines are the observable expectations of the real system as functions of the evolution time, and the scatters are the predicted values given by the digital twin.
  (c) Sensing precision of the different schemes, described by the CFI in the measurement base $\sigma_z$. 
  (d) Evolution trajectories of the system with different schemes, represented by the observable expectation.
  The inset show the control actions of the different protocols.}
  \label{case1fig}
\end{figure}

\textit{Single atom.}---We consider an atom driven by a classical field with the Hamiltonian
\begin{equation}
{H_S} = {\sigma _z}\sin 2\theta  + {\sigma _x}\cos 2\theta,
 \label{atom}
\end{equation}
where $\theta$ is the direction of the classical field to be estimated.
In the noise-free model, since the Pauli terms of the encoding parameters are not commute, additional control is required to achieve the HL. 
According to Ref.~\cite{olc1}, the optimal control is $\sigma_z$. 
Considering open system, we model the coupling term between the environment and the system as ${H_I} = g{\bm{\sigma} }({b^+}b)$, where $\bm{\sigma}$ is the noise operator of system and we set $\bm{\sigma}=H_S$.
In our protocol, the environment needs to be continuously monitored. 
Thus the evolution of the system and the environment is described by the stochastic master equation \cite{sme1,sme2}
\begin{equation}
\begin{gathered}
  {{\dot \rho }_{SE}} =  - i[{H_{SE}},{\rho _{SE}}] \hfill \\
   \;\;\;\;\;\;\;+ \kappa {\cal D}[A]{\rho _{SE}} + \sqrt {\kappa \eta } {\cal H}[A]{\rho _{SE}}dW, \hfill \\ 
\end{gathered}
 \label{atomsme}
\end{equation}
where $\kappa$ is the measurement rate, $\eta$ is the detection efficiency, $A$ is measurement operator, ${\cal D}[ \cdot ]$ or ${\cal H}[ \cdot ]$ is superoperator describing respectively, and $dW$ is random Wiener increment (see SM).
{1.00,0.00,0.00}{This equation can be simulated numerically using the QuTiP library \cite{qutip}.}
In our framework, the continuous monitoring environment maintains the purity of the system. 
Then the RL can learn an adaptive compensation strategy that eliminates the random measurement backaction under different trajectories. 
{For the system with continuous measurements, if the measurement rate is sufficiently high and perfect detection ($\eta=1$) is achieved, the quantum state remains purity under each measurement trajectory (see SM).
In Fig.~\ref{case1fig}(a), we show the system dynamics under the measurement setting ($\sqrt{\kappa}=1.5,\eta=1$).}
The purity of the system remains constant, but the variance of the state under different trajectories grows rapidly with time.
{Throughout the main text, we assume perfect detection is achievable.}

Here we use $\{ \left\langle {{\sigma _x}} \right\rangle ,\left\langle {{\sigma _y}} \right\rangle ,\left\langle {{\sigma _z}} \right\rangle \}$ as the states of the digital twin model. 
In the RL phase, we use the control Hamiltonian of the form ${H_C} = {c_x}{\sigma _x} + {c_y}{\sigma _y} + {c_z}{\sigma _z}$. 
Considering experimental convenience, we use CFI as the reward function. 
In Fig.~\ref{case1fig}(b), we demonstrate the ability of the digital twin model to track the real system, which accurately reﬂects the evolution of the real system. 
In Fig.~\ref{case1fig}(c), we compare the sensing precision of the different protocols in noisy environment. 
For the noise-free case, the highest precision that can be achieved is the HL.
In noisy environment, the precision with non-control (NC) is almost 0, and the precision of the open-loop control (OLC) \cite{olc1} protocol is below the SQL. 
In comparison, our protocol restores the HL. 
{Since QEC protocols cannot guarantee HL recovery when the noise operator commutes to the Hamiltonian \cite{zss1}, our protocol overcomes this limitation. In the SM, we detail analyze of the conditions of HL for our protocol.}
{Moreover, the effectiveness of QDTS relies on perfect detection to ensure the system remains in a pure state. Imperfect detection or delayed feedback introduces noise into the digital twin model. In the SM, we analyze the impact of these factors on QDTS and observe its robustness.}


Finally, we analyze how the QDTS protocol reaches the HL under environmental noise. Fig.~\ref{case1fig}(d) compares the quantum system evolution trajectories for different control schemes. The QDTS protocol’s adaptive compensation strategy significantly suppresses environmental noise-induced trajectory diffusion while maintaining an average trajectory nearly identical to the noise-free case. This indicates that QDTS effectively emulates noise-free dynamics, achieving performance comparable to QEC without requiring additional Hilbert space resources. In contrast, open-loop control fails to mitigate trajectory diffusion, resulting in a mean trajectory that substantially deviates from the noise-free evolution. This discrepancy not only increases estimation variance but also introduces systematic bias in parameter inference \cite{inference1,bias1}. Remarkably, our protocol unexpectedly mitigates this noise-induced bias while preserving the HL scaling of estimation precision.

\begin{figure}[h]
  \centering
  \includegraphics[width=0.48\textwidth]{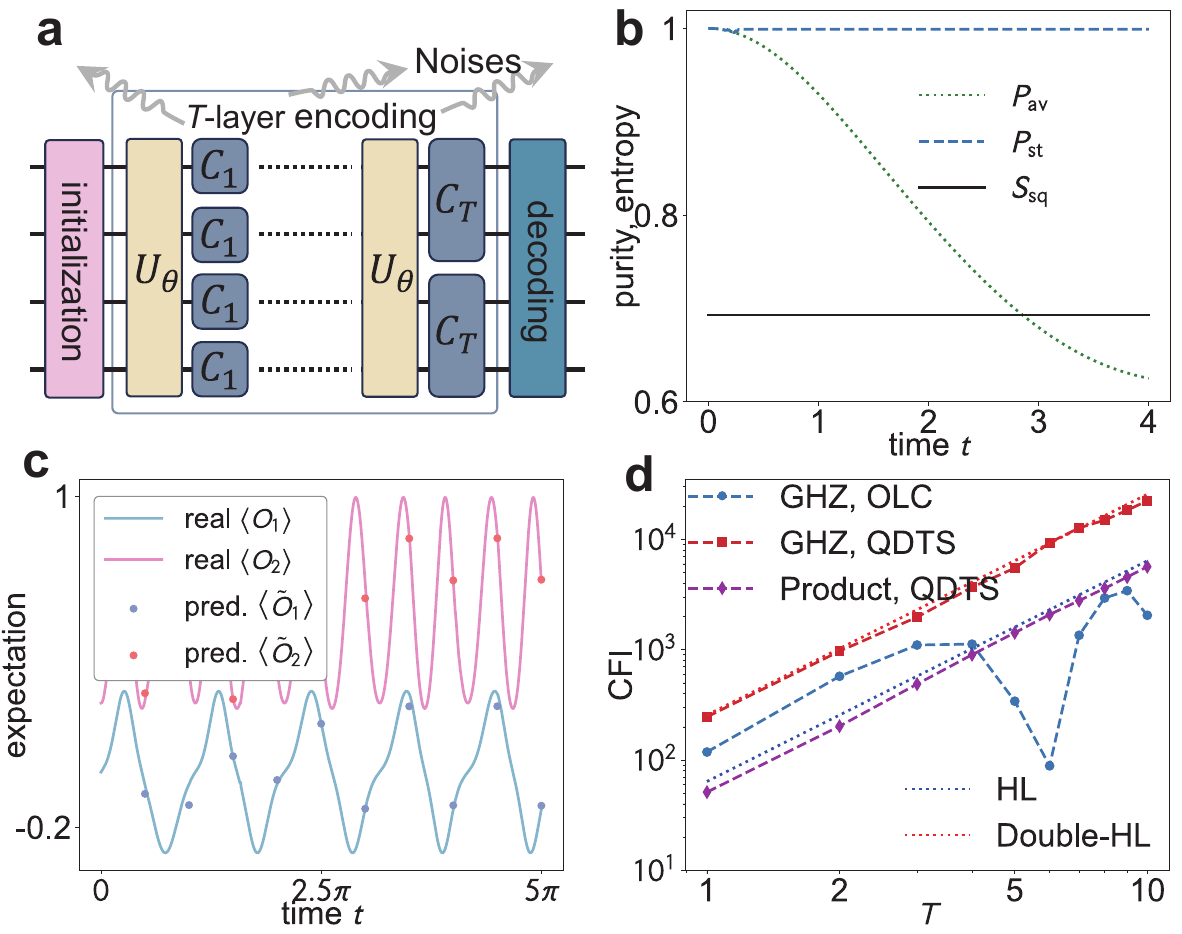}
  \caption{Results for the quantum circuit system.
  (a) Schematic of the quantum sensing circuit.
  (b) Purity and entropy of circuit for GHZ state case under the continuous monitoring environment setup.
  $P_{\rm st}$ is the purity of the density matrix of a single trajectory, while $P_{\rm av}$ is the purity after averaging the density matrix over all trajectories.
  (c) Predictive performance of the digital twin model. 
  (d) Sensing precision of the different schemes, described by the CFI in the optimal measurement basis.}
  \label{case2fig}
\end{figure}

\textit{Quantum circuit.}---We study an $n$-qubit quantum circuit sensing with the circuit structure depicted in Fig.~\ref{case2fig}(a). 
The circuit comprises three stages: (i) initialization, where the system is prepared in either a product state or a GHZ state; (ii) parameter encoding, implemented via $T$ sequential circuit layers, each containing an unknown single-qubit rotation gate followed by a control gate (single- or two-qubit local gates); and (iii) decoding, which realizes the optimal measurement basis. 
Furthermore, the entire circuit is coupled with the environment, resulting in decoherence.
We take the example of a 4-qubit circuit.
The unknown parameter gate is
\begin{equation}
{U_\theta } = \prod\limits_{j = 0}^{n=4} {{e^{i\pi ({\sigma _z^j}\sin 2\theta  + {\sigma _x^j}\cos 2\theta )/2}}}.
 \label{gate}
\end{equation}


{For this model, using the density matrix in the QDTS protocol is not scalable due to the tomographic requirement of $4^{n}-1$ observables. 
However, when initialized to a product state, the digital twin model only needs 3 observables $\{ \left\langle {\sum {\sigma _x^i} } \right\rangle ,\left\langle {\sum {\sigma _y^i} } \right\rangle ,\left\langle {\sum {\sigma _z^i} } \right\rangle \}$. 
For the GHZ state, only 9 observables are required (see SM).
As shown in Fig.~\ref{case2fig}(b), under continuous monitoring, the GHZ state maintains its purity, while ignoring measurements leads to a mixed state. 
For a single trajectory, the single-qubit entropy remains at the maximum value $S_{\rm sq}=\ln 2$, indicating maximal entanglement. 
Thus, the system stays within the GHZ subspace, and a small number of observables suffice to capture its evolution.}
In Fig.~\ref{case2fig}(c), we show the prediction performance of the digital twin model for GHZ state case. It faithfully tracks the evolution of the real system.

In Fig.~\ref{case2fig}(d), we present the precision limits attained by the QDTS protocol for both product and GHZ states under noisy environment. 
For product state case, the QDTS restores the HL ${{\cal F}_C} \propto n{T^2}$. 
Remarkably, for GHZ state case, the precision reaches the double HL ${{\cal F}_C} \propto {n^2}{T^2}$. 
In contrast, the OLC protocol fails to surpass the SQL.
The results demonstrate that QDTS effectively learns the stochastic errors during  decoherence process of quantum circuit. 
Then by adaptively compensating for these errors, it suppresses noise effects on both coherence and entanglement.

\textit{Continuous variable.}---Here we consider a sensing task for a single bosonic mode with the Hamiltonian
\begin{equation}
{H_S} = {\omega _0}{a^ + }a + \chi (\xi {a^2} + {\xi ^*}{({a^ + })^2}),
\label{optic}
\end{equation}
where $\omega_0$ is the unknown parameter, $\chi$ is the squeezing strength, and $\xi$ is related to the squeezing direction.
We first analyze the noise-free case.
If squeezing is not available, i.e., $\chi = 0$, the precision of estimating $\omega_0$ can be realized in the HL. 
If squeezing is available, the super-HL (SHL) can be achieved, and its precision beyond the HL stems from the squeezing accumulated in the evolution. 
However, we consider open system. The coupling term between the system and the environment is modeled as $H_I=g{a^ + }a{b^ + }b$. 
In the QDTS protocol, we use $\{ \left\langle p \right\rangle ,\left\langle q \right\rangle ,{\left\langle p ^2\right\rangle},{\left\langle q^2 \right\rangle},\left\langle {pq} \right\rangle \}$ as state of digital twin model where $p$ and $q$ are the  quadrature operators of the system.
These features are sufficient to characterize the real system in both the non-squeezing and squeezing cases (see SM). 
In RL, the CFI on the homodyne measurement basis is used as the reward function, and the control Hamiltonian is ${\omega}{a^ + }a$.

\begin{figure}[h]
  \centering
  \includegraphics[width=1\linewidth]{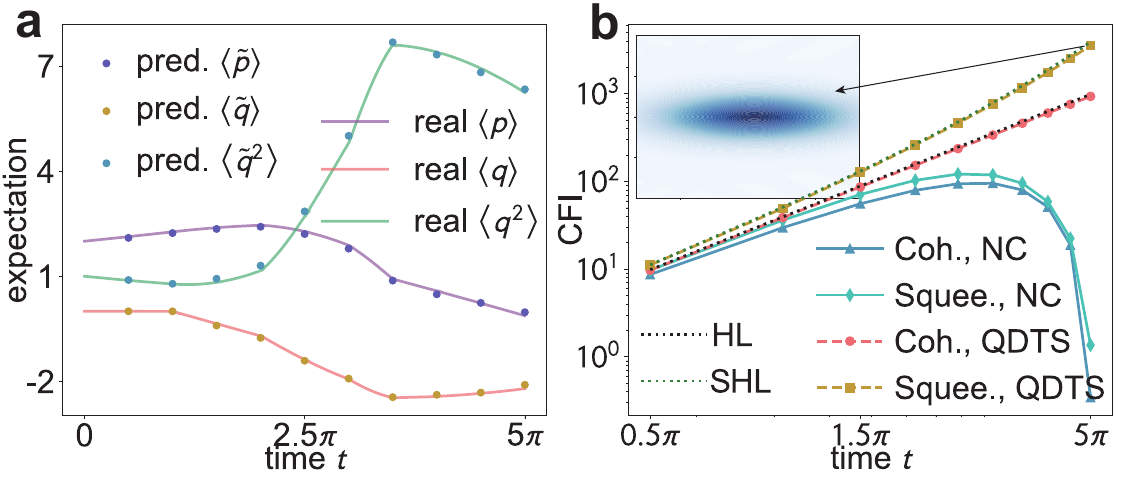}
  \caption{Results for the continuous variable system.
  (a) Predictive performance of the digital twin model. 
  (b) The sensing precision of all considered protocols is quantified via the CFI, evaluated through homodyne measurements.
  The inset shows the final state of the system using QDTS protocol in the squeezing case.}
  \label{case3fig}
\end{figure}

In Fig.~\ref{case3fig}(a), we show the tracking performance of the digital twin  model for a real system, where the predicted observations can accurately reflect the evolution of the system. 
In Fig.~\ref{case3fig}(b), we show the curves of the CFI on the homodyne measurement basis of the QDTS protocol in noisy environment. 
They both restore the HL or super-HL for the coherent (non-squeezing) or squeezing case. 
In contrast, the sensing precision without control (NC) degrades rapidly beyond the coherence time, both in the coherent and squeezing cases. 
Additionally, the inset shows the final state of the system when QDST is employed, which is the same as the final state in the absence of noise (see SM).
These results show that QDTS also learns real-time noise compensation in this model, eliminating the effect of noise on squeezing.

\textit{Conclusions.}---We propose the QDTS protocol and demonstrate its efficacy across diverse physical systems. 
Designed for noise-agnostic scenarios, QDTS provides a control framework to restore the HL.
Specifically, QDTS first learns the dynamics of the real system from continuous measurement data and builds its digital twin model. 
Then, it learns the optimal control strategy based on the features predicted by the twin system. 
By extracting real-time information about random errors in the decoherence process by continuous measurements, QDTS realizes dynamic error compensation. 
In single atom, quantum circuit, and continuous variable systems, QDTS demonstrates its ability to eliminate the effects of noise on coherence, entanglement, and squeezing, restoring HL achievable without noise.
These results overturn the long-held belief that the unitary control cannot restore HL for noisy quantum sensing.
{Notably, QDTS is applicable in scenarios without auxiliary qubits or prior knowledge of the noise operator. 
It restores the Heisenberg limit in noisy channels where QEC protocols are infeasible, thereby complementing the limitation of QEC.}
Additionally, QDTS is not only applicable to various noisy sensing scenarios, but also provides a new noise suppression framework for physics problems in the NISQ era, 
such as quantum battery charging \cite{battery1,battery2}, ground state preparation \cite{ground1,ground2}, error correction codes \cite{correction1,correction2} and entanglement preservation for quantum communication \cite{qc1}.

\textit{Note added.}---As pointed out by Uwe Fischer, a recent work \cite{kwon2025restoring} proposes an AutoQEC protocol to restore the Heisenberg limit in time. We thank him for bringing this to our attention.

\textit{Acknowledgments.}---This work was supported by the National Natural Science Foundation of China (No.~62401359), the fund of State Key Laboratory of Photonics and Communications, the Innovation Program for Quantum Science and Technology (No.~2021ZD0300703), Shanghai Municipal Science and Technology Major Project (No.~2019SHZDZX01) and SJTU-Lenovo Collaboration Project (No.~202407SJTU01-LR019).

\nocite{*}

\bibliography{sample}

\end{document}